\documentclass[a4paper,11pt]{article}
\pdfoutput=1 % if your are submitting a pdflatex (i.e. if you have
\usepackage{jheppub} % for details on the use of the package, please
\usepackage[T1]{fontenc} % if needed
\usepackage{adjustbox}
\usepackage{comment}
\usepackage{soul}
\DeclareMathOperator{\Tr}{Tr}
\def\Re{\text{Re}}

\def\cK{{\cal K}}
\def\cJ{{\cal J}}
\def\cP{{\cal P}}

\title{A modified in-medium evolution equation with color coherence}

\author[a]{Jo\~ao Barata}
\author[a]{ Fabio Dom\'inguez}
\author[a]{ Carlos A. Salgado}
\author[b]{ V\'ictor Vila}

\affiliation[a]{Instituto Galego de F\'isica de Altas Enerx\'ias IGFAE, Universidade de Santiago de Compostela,
E-15782 Galicia-Spain}

\affiliation[b]{CPHT, CNRS, École Polytechnique, Institut Polytechnique de Paris, 91128 Palaiseau, France}

\emailAdd{joaolourenco.henriques@usc.es}
\emailAdd{fabio.dominguez@usc.es}
\emailAdd{carlos.salgado@usc.es}
\emailAdd{victor.vila@polytechnique.edu}

\abstract{
QCD jets produced in heavy-ion collisions at LHC or RHIC energies partially evolve inside the produced hot and dense quark gluon plasma, offering unique opportunities to study QCD splitting processes in different backgrounds. Induced (modified) splittings are expected to be the most relevant mechanism driving the modifications of in-medium jets compared to vacuum jets for a wide sets of observables. Although color coherence among different emitters has been identified as an essential mechanism in studies of the {\it QCD antenna radiation}, it is usually neglected in the multi-gluon medium-induced cascade. This independent gluon emission approximation can be analytically proved to be valid in the limit of very large media, but corrections or modifications to it have not been computed before in the context of the evolution (or rate) equation describing the gluon cascade. We propose a modified evolution equation that includes corrections due to the interference of subsequent emitters. In order to do so, we first compute a modified splitting kernel following the usual procedure of factorizing it from the subsequent Brownian motion. The calculation is performed in the two-gluon configuration with no overlapping formation times, that is expected to provide the first correction to the completely independent picture.
}

\begin{document} 
\maketitle
\flushbottom

\section{Introduction}\label{sec:intro}
One of the strongest evidences for the creation of the Quark Gluon Plasma (QGP) at RHIC~\cite{RHIC1,RHIC2} and LHC~\cite{LHC1,LHC2,LHC4,LHC5} is jet quenching: the modification of jets due to the interaction with the dense QCD medium created in high-energy collisions of heavy atomic nuclei. The most direct observable consequence of this effect is the suppression of the yields of particles and jets at large transverse momentum --- the quenching. However, {\it jet quenching} is nowadays a generic name that embraces the modern technology of jet studies, originally developed for jets in vacuum (i.e. in proton-proton or simpler colliding systems), including a plethora of global or sub-jet observables with different degrees of sophistication. These new observables pose a challenge on present theoretical descriptions of in-medium jet cascades that are stimulating advances towards a more precise implementation of the underlying physics. 

Jets in heavy-ion collisions develop partly inside the surrounding QCD matter and partly outside of it, with quantum interference between the two possibilities. Moreover, the total shower contains both medium-induced radiation as well as (angular-ordered, infrared and soft divergent) vacuum contributions. One of the main theoretical difficulties to write a consistent cascade is to understand how to order the subsequent splittings of the two kinds. Under some circumstances, for which color coherence between the different emitters in the cascade plays a central role, the vacuum and medium contributions to the cascade can be factorized \cite{Antenna4,Caucal}. Before a more complete description is available, a usual approximation is to evolve both cascades independently. For soft gluons, which in the medium have small formation time $t_{f}$, it can be shown that interference between subsequent emitters can be neglected for large enough media, $t_{f}/L\ll 1$ \cite{BDIM1,Liliana2}. This independence ensures a probabilistic picture in which evolution equations (known as {\it rate equations}) can be easily computed for different jet properties \cite{BDIM2,Jeon:2003gi}. The goal of the present paper is to go beyond this approximation, taking into account the first correction to the completely independent subsequent gluon emission and to propose a modification of the rate equations that takes into account color coherence. 

The single gluon production, the building block for the in-medium cascade, has been extensively studied along the past few decades~\cite{BDMPS1,BDMPS2, BDMPS3,BDMPS4,GLV,Wiedemann,Wang:2001ifa,Arnold:2002ja,BDIM1,Liliana2,Sievert:2019cwq} and although full numerical solutions are well established~\cite{numerical1, numerical2, numerical3}, a fully analytic formulation has not been achieved\footnote{See~\cite{CarlotaFabioLiliana,CarlotaFabioMarcos,IOE1,IOE2,IOE3} for recent efforts.}. Nonetheless, for sufficiently large and dense media, the spectrum is well described by the Baier-Dokshitzer-Mueller-Peigné-Schiff-Zakharov (BDMPS-Z) framework~\cite{BDMPS1,BDMPS2, BDMPS3,BDMPS4}, which encapsulates the propagation of an energetic parton that exchanges multiple soft gluons with the medium. In this regime, the main mechanism for energy loss consists in the emission of induced soft radiation with frequency $\omega$ above the Bethe-Heitler bound, $\omega\gg\omega_{\rm BH}\sim \frac{\mu^4}{\hat{q}}$, but below the critical frequency, $\omega\ll\omega_c\sim\hat{q}L^2$, with a typical formation time $t_f(\omega)=2\omega/\mathbf{k}^2$, where $\mathbf{k}$ is the transverse momentum of the gluon, $L$ the medium length, $\mu$ the Debye screening mass and $\hat{q}$ the averaged square transverse momentum acquired by a particle propagating in the medium during a time t, i.e. $\langle \mathbf{k}^2\rangle= \hat{q}t$. Using the previous estimates, one has that $t_f\sim \omega/(\hat{q}t_f)\sim \sqrt{\omega/\hat{q}}$, with the gluon acquiring a transverse momentum $\mathbf{k}^2\sim \hat{q}t_f\sim \sqrt{\hat{q}\omega}$ during the branching process. In the limit where multiple gluon emissions are observed (i.e. $\omega\ll\omega_c$\footnote{See the discussion on the multiple soft emission region conducted in~\cite{BDIM1,BDIM2}.}), we have that gluon radiation is formed almost instantly since $t_f(\omega)\ll t_f(\omega_c)=L$, while the final transverse momentum of the gluon $\mathbf{k}^2\sim \hat{q}L\gg \sqrt{\hat{q}\omega}\gg\mu^2$, as the gluon still has to propagate until the end of the medium after its formation. Thus, to leading order in inverse powers of the medium length, soft gluons are produced decoherently and almost instantaneously, and the probability to emit a gluon is proportional to $L-t_f(\omega)\approx L$.\par 
This discussion can also be formulated in terms of the angular structure of the emission spectrum. Defining the emission angle $\theta^2=\frac{\mathbf{k}^2}{\omega^2}$, we have that $\theta^2\sim \frac{1}{\hat{q}t_f^3(\omega)}\gg\frac{1}{\hat{q}L^3}\equiv \theta_c^2$; in contrast, the measured angle gets its main contribution from final state broadening, as can be verified by noticing that accumulated transverse momentum is proportional to the traversed length $L\gg t_f$, while the energy is conserved. This justifies a picture of time localized splittings (as $t_f\ll L$) producing decoherent partons, with the overall transverse structure being determined by individual momentum broadening of the final states~\cite{BDIM1,Antenna2}.

More recently, a lot of effort has been put into studying multiparticle interference effects absent from the BDMPS-Z picture. One of such effects is color coherence between emitters, that has been considered in studies exploring the physics of the QCD antenna with an extra in-medium gluon emission~\cite{Antenna0,Antenna1,Antenna2,Antenna3,Antenna4}. The main conclusions of such studies were that for short time scales and emission angles, partons keep color coherent and splittings are not immediately resolved by the medium, while for long time intervals or large emission angles, the medium randomizes the color fields of each parton such that the system evolves decoherently. Generalizing such a picture for a full in-medium jet~\cite{Antenna4}, it was argued that the color coherence between emitters within the in-medium shower might lead to a significant modification of the expected gluon spectrum for relevant experimental conditions. In addition, it is also well-known that in vacuum, color coherence between emitters needs to be taken into account in order to properly describe experimental data~\cite{Azimov:1985by}.\par

In this paper we include, for the first time, color coherence effects in the resummation of multiple gluon emissions. 
In particular, color coherence is included by allowing partons to take a finite time to be resolved by the medium after the splitting, followed by decoherent final state broadening\footnote{See~\cite{Hulcher:2017cpt} for a qualitative similar idea introduced within the context of the Hybrid model~\cite{Hybrid}.}. Since splittings are still sharply localized when compared to the scale $L$, we can take the single gluon branching process as the building block for a probabilistic gluon shower, similarly to the totally decoherent case~\cite{BDIM2,bottom_up_therm}.
The introduction of color coherence however leads the final evolution equation for the gluonic shower to become, in general, non-local in time, since it takes a finite amount of time for partons to color decohere. The details of the color coherence dynamics are obtained by studying the emission of a soft gluon (the same set up as BDMPS-Z) followed by the emission of a vacuum soft gluon which serves as a probe of the color coherence of the two outgoing states. An effective 
coherence factor is then extracted and applied as a correction to the emission kernel used in the totally decoherent case. Although this \textit{ansatz} approach does not strictly follow directly from a first principle calculation, it allows us to gauge the effects of including color coherence effects at the level of each splitting.

The present paper is divided as follows. Section \ref{section:set_up} gives an overview of previous works on the resumation of multiple decoherent in-medium gluon emissions~\cite{BDIM1,BDIM2,BDIM3}; Section~\ref{sec:computing} presents the computation of the interference due to including a soft vacuum emission, while Section~\ref{sec:evolution_eqs} provides the derivation of the new evolution equation for the gluon shower. The conclusions are presented in Section~\ref{sec:conclusion}. Further details are provided in three appendices.

\section{Theoretical Set Up}\label{section:set_up}
In this section we briefly review the main results from~\cite{BDIM1,BDIM2}, where the double-differential medium-induced gluon emission spectrum was computed and simplified in order to provide a probabilistic picture for the production of medium-induced radiation\footnote{See~\cite{Liliana2} for related work.}.

\subsection{Gluon emission spectrum off a parton}\label{subsec:BDMPS_full}
When an energetic parton propagates in a dense QCD medium, it exchanges multiple soft gluons with the medium. The leading order effect of such interactions is to transversely kick the hard parton. The probability $\mathcal{P}_1(\mathbf{k};t,t_0)$  that the parton acquires a transverse momentum $|\mathbf{k}|\ll p_0^+$, with $p_0^+$ the parton energy, due to such interactions during a time $L-t_0$  is given by (see \cite{BDIM1,BDIM2} and references therein)  
\begin{equation}
\mathcal{P}_1(\mathbf{k};L,t_0)=\int_\mathbf{r} e^{-i \mathbf{r}\cdot \mathbf{k}}   \mathcal{P}(\mathbf{r};L,t_0)=\int_\mathbf{r} e^{-i \mathbf{r}\cdot \mathbf{k}}   \, e^{-\frac{C_A}{2}\int_{t_0}^Ldt \, n(t) \sigma(\mathbf{r})} \, ,
\end{equation}
where $\mathcal{P}(\mathbf{r})$ is the dipole operator in the adjoint representation, $n=n(t)$ the density of scattering centers, and $\sigma$ the in-medium elastic cross-section (see Appendix \ref{append:averages}). Taking the derivative with respect to $L$ one obtains the following evolution equation for this probability distribution,
\begin{equation}\label{eq:derivative_P1}
\partial_L \mathcal{P}_1(\mathbf{k};L,t_0)=\int_\mathbf{l} \mathcal{C}(\mathbf{l},L)\mathcal{P}_1(\mathbf{k}-\mathbf{l};L,t_0)   \, , 
\end{equation}
where for the momentum space integrals we use the shorthand $\int_\mathbf{q}=\int (2\pi)^{-2} d^2\mathbf{q}$, and for the position space integrals we use $\int_\mathbf{r}=\int d^2\mathbf{r}$. Here the broadening kernel is given by  
\begin{equation}
\mathcal{C}(\mathbf{l},t)=-\frac{C_A}{2}n(t)\sigma(\mathbf{l})   \, .
\end{equation}
%where $n=n(t)$ is the density of scattering centers and $\sigma$ the in-medium elastic cross-section. In this paper we will typically assume $n$ to be time independent.\par

\begin{figure}[h!]
    \centering
    \includegraphics[scale=0.6]{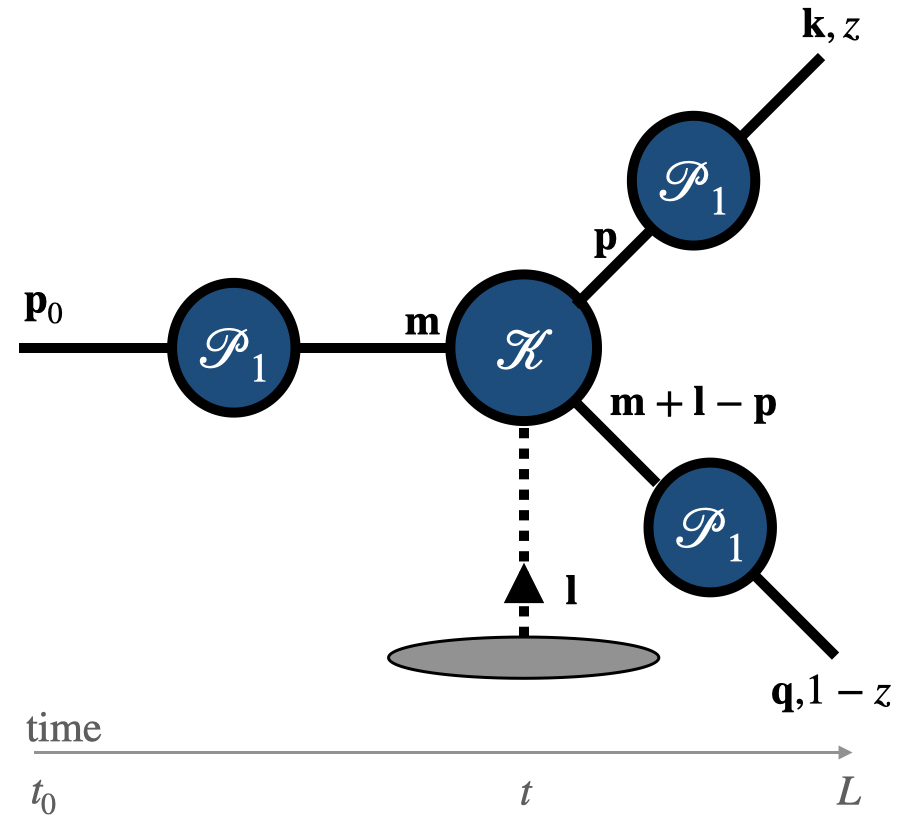}
    \caption{Diagrammatic representation of eq. \eqref{eq:general_P2}. The labels used follow the notation in the main text and the gray blob denotes the underlying QCD medium. }
    \label{fig:P2_fig}
\end{figure}

In addition to momentum broadening, the multiple interactions with the medium also induce the production of soft radiation. In a similar fashion, one can construct the probability $\cP_2(\mathbf{k},\mathbf{q};L,t_0)$ of observing two outgoing partons, with transverse momentum $\mathbf{k}$ and $\mathbf{q}$ respectively, from an initial state with momentum $\overrightarrow{p_0}=(p_0^+,\mathbf{p}_0)$, for a process happening between times $t_0$ and $L$. Taking into account that the soft modes have typical formation times inside the medium much smaller than the medium length $t_f\ll L$, one can effectively ignore the formation time of radiation when compared to any other time scale\footnote{See \cite{BDIM1,BDIM2} for a detailed discussion of this approximation, in a notation close to the one implemented in this manuscript. More details can also be found in the references therein and follow the qualitative discussion in the previous section.}. In this approximation, the two outgoing states evolve decoherently at late times and one can write~\cite{BDIM1,BDIM2}

\begin{equation}\label{eq:general_P2}
\begin{split}
\cP_2(\mathbf{k},\mathbf{q},z;L,t_0)&=2g^2z(1-z)\int_{t_0}
^L dt \, \int_{\mathbf{m},\mathbf{Q},\mathbf{l}} \cK(\mathbf{Q},\mathbf{l},z,p_0^+;t)
\\
&\times\cP_1(\mathbf{m}-\mathbf{p}_0;t,t_0)\cP_1(\mathbf{k}-\mathbf{p};L,t)\cP_1(\mathbf{q}-(\mathbf{m}+\mathbf{l}-\mathbf{p});L,t) \, .    
\end{split}
\end{equation}
Here $\mathbf{l}$ is the transverse momentum acquired during the branching process, $\mathbf{m}$ the momentum of the initial parton before splitting, $\mathbf{p}$ the transverse momentum of the outgoing parton with energy $zp_0^+$ just after the splitting, and $\mathbf{Q}=\mathbf{p}-z(\mathbf{m}+\mathbf{l})$ the relative transverse momentum of the system after branching -- see Figure \ref{fig:P2_fig}.

Noticing that $\mathbf{Q}$ and $\mathbf{l}$ are the only momenta scales directly entering the branching process, they can be neglected in a regime that makes it possible to build a probabilistic picture for the emission process. The observation that such a region exists can be argued as follows. \par 
The scale $\mathbf{l}$ is generated by transverse momentum broadening during the branching process and thus $\mathbf{l}^2\sim \hat{q}t_f\ll \hat{q}L$, which can be neglected with respect to $\mathbf{k}^2\sim\mathbf{q}^2\sim \hat{q}L$. Then, disregarding such a scale in the single particle broadening contributions and integrating $\cK$ over $\mathbf{l}$ we get \par

\begin{equation}\label{eq:P2_wQ}
\begin{split}
\cP_2(\mathbf{k},\mathbf{q},z;L,t_0)&=2g^2z(1-z)\int_{t_0}
^L dt \, \int_{\mathbf{m},\mathbf{Q}} \cK(\mathbf{Q},z,p_0^+;t)
\\
&\times\cP_1(\mathbf{m}-\mathbf{p}_0;t,t_0)\cP_1(\mathbf{k}-\mathbf{p};L,t)\cP_1(\mathbf{q}-(\mathbf{m}-\mathbf{p});L,t) \, ,    
\end{split}
\end{equation}
with $\mathbf{Q}=\mathbf{p}-z\mathbf{m}$. 

The relative momentum $\mathbf{Q}$ is a purely kinematical scale, i.e. it captures how non-collinear the outgoing parton's momentum are after the branching. Therefore, in this sense, there is \textit{a priori} no constraint on the values it can take. However, the magnitude of $\mathbf{Q}$ is determined by the BDMPS-Z splitting kernel $\cK$, which, as we will show below, is peaked around $\mathbf{Q}^2\sim\sqrt{\hat{q}zp_0^+}\ll \hat{q}L$ (for small $z$), with smaller momentum scales blocked by the LPM coherence effect and larger values being exponentially harder to obtain via multiple soft scattering~\cite{BDIM1,BDIM2}. As a consequence, one can further simplify the above relation to

\begin{equation}\label{eq:P2}
\begin{split}
\cP_2(\mathbf{k},\mathbf{q},z;L,t_0)&=2g^2z(1-z)\int_{t_0}
^L dt \, \int_{\mathbf{m},\mathbf{Q}} \cK(\mathbf{Q},z,p_0^+;t)
\\
&\times\cP_1(\mathbf{m}-\mathbf{p}_0;t,t_0)\cP_1(\mathbf{k}-z\mathbf{m};L,t)\cP_1(\mathbf{q}-(1-z)\mathbf{m};L,t)   \\
&\equiv2g^2z(1-z)\int_{t_0}
^L dt \, \int_{\mathbf{m}} \cK(z,p_0^+;t)
\\
&\times\cP_1(\mathbf{m}-\mathbf{p}_0;t,t_0)\cP_1(\mathbf{k}-z\mathbf{m};L,t)\cP_1(\mathbf{q}-(1-z)\mathbf{m};L,t)  \, ,
\end{split}
\end{equation}
where we used the fact that neglecting the momentum exchanges during branching leads to a collinear splitting. The splitting kernel is given by~\cite{BDIM2}
\begin{equation}\label{eq:K_energy}
\mathcal{K}(z,p_0^+;t)=\frac{P_{gg}(z)}{2\pi}\sqrt{\frac{\hat{q}(t)(1-z+z^2)}{p_0^+ z(1-z)}} \, ,
\end{equation}
where purely gluonic degrees of freedom are assumed. Here $P_{gg}$ is the Altarelli-Parisi vacuum kernel multiplied by a factor of $C_A$ and the time dependence can be dropped as long as one assumes that the medium is static and homogeneous (plasma brick model).\par

\subsection{The Generating Functional and the shower building blocks}\label{subsec:functional_ev}
Using the results from the previous section, we now derive an evolution equation for the single gluon inclusive distribution resumming multiple medium-induced gluon emissions. We make use of the generating functional method~\cite{Book1,Book2,Pedrag}, although this is
not crucial.\par 
We first consider the functional $\mathcal{Z}_{p_0}(u;t,t_0)$ within the time interval  $t_0\leq t\leq L$,
\begin{equation}\label{eq:Z_functional}
\mathcal{Z}_{p_0}(u;t,t_0)=\sum_{n=1}^\infty \frac{1}{n!} \int_{\Omega_n}P_n(\overrightarrow{k_1},\ldots,\overrightarrow{k_n};t,t_0)u(\overrightarrow{k_1})\ldots u(\overrightarrow{k_n})    \, ,
\end{equation}
where the integration is performed over all the individual phase-spaces $\Omega_n$ on each term of the sum, and $u(\overrightarrow{k})$ is a test function that will eventually drop out via functional differentiation\footnote{We define the functional derivative as $\frac{\delta u\left(\overrightarrow{p}\right)}{\delta u(\overrightarrow{q})}=\delta(p^+-q^+)\delta(\mathbf{p}-\mathbf{q})$. Inclusive distributions (as we are interested in computing) can be obtained from functional $\mathcal{Z} $ by taking the functional derivative with respect to $u$ at $u=1$~\cite{Pedrag}.}. All possible physical processes are stored in $\mathcal{Z}$ via the elementary probabilities $P_n(\overrightarrow{k_1}, \ldots, \overrightarrow{k_n})$, which correspond to the probability of measuring $n$ final-state gluons with the momentum assigned to each one at time $t$. For the case at hand, the evolution is defined by the single particle broadening probability $P_1$ and the branching probability $P_2$. These are related to the probabilities $\mathcal{P}_n$ introduced in the previous section by

\begin{equation}
P_n(\overrightarrow{k_1},\ldots,\overrightarrow{k_n};t,t_0)=2p_0^+(2\pi) \delta\left(\sum_{i=1}^n k^+_i-p_0^+\right) \mathcal{P}_n(\mathbf{k}_1,\ldots,\mathbf{k_n};t,t_0) \, , 
\end{equation}
where the energy fractions $z_i$ have been omitted for the sake of clarity. This relation makes it explicit that the dynamics are constrained to the transverse plane, since $p_0^+=\sum_{i=1}^{n}k_i^+ $ and the remaining freedom in the $k_i^+$ is fixed by the splitting energy fractions $z_i$.\par

In order to obtain an evolution equation for the functional, an evolution law for $\mathcal{P}_2$ is needed. However, as this probability already includes broadening contributions associated to in/out-going legs (see Figure \ref{fig:P2_fig}), we truncate such terms by introducing the associated branching probability 
\begin{equation}\label{eq:P2tilde}
\begin{split}
\widetilde{\mathcal{P}}_2(\mathbf{k},\mathbf{q},z;L,t_0) &= 2g^2z(1-z)\int_{t_0}^{L}\  dt \  \mathcal{K}(z,p_0^+;t) (2\pi)^4\delta^{(2)}(\mathbf{k}-z\mathbf{p}_0)\delta^{(2)}(\mathbf{q}-(1-z)\mathbf{p}_0)
\end{split} \, ,
\end{equation}
where we have already performed the integration over the initial delta function. Notice that by doing this, double counting contributions already included in $P_1$ is avoided. The time-evolution equation is then given by
\begin{equation}\label{eq:evol_P2_sec1}
\begin{split}
\partial_L \widetilde{\mathcal{P}}_2(\mathbf{k},\mathbf{q},z;L,t_0)=2 g^2z(1-z) \mathcal{K}(z,p_0^+;L)(2\pi)^4\delta^{(2)}(\mathbf{k}-z\mathbf{p}_0)\delta^{(2)}(\mathbf{q}-(1-z)\mathbf{p}_0) \, .
\end{split}
\end{equation}
Now, combining this result with the time-evolution equation for $\mathcal{P}_1$ and taking into account that for an infinitesimal time step $dt$
\begin{equation}\label{eq:final1}
\begin{split}
\mathcal{Z}_{p_0}(t_0+dt,t_0)&=  \int d\Omega_k P_1(\overrightarrow{k},t_0+dt,t_0) u(\overrightarrow{k}) 
\\&+\frac{1}{2}\int d\Omega_{k_1}d\Omega_{k_2} P_2(\overrightarrow{k}_1,\overrightarrow{k}_2;t_0+dt,t_0)u(\overrightarrow{k_1})u(\overrightarrow{k_2}) \, ,
\end{split}
\end{equation}
one can eventually write the evolution law for the functional 
\begin{equation}\label{eq:Z_t_evolution}
\begin{split}
\partial_t \mathcal{Z}_{p_0}(t,t_0| u)\bigg|_{t=t_0}&=\int_\mathbf{l} C(\mathbf{l},t) u(p_0^+,\mathbf{p}_0+\mathbf{l})
\\&+\alpha_s\int_z \mathcal{K}(z,p_0^+;t)\left[u(z \overrightarrow{p_0})u((1-z)\overrightarrow{p_0})-u(\overrightarrow{p_0})\right] \, ,
\end{split}
\end{equation}
where the last term in the $\mathcal{O}(\alpha_s)$ bracket comes from probability conservation (so that when $u=1$ the right-hand side vanishes).  This relation can be extended to the full shower
\begin{equation}\label{eq:Z_t_evolution_1}
\begin{split}
\partial_t \mathcal{Z}_{p_0}(t,t_0| u)&=\int_{q^+\mathbf{q}\mathbf{l}} C(\mathbf{l},t) u(q^+,\mathbf{q}+\mathbf{l}) \frac{\delta  \mathcal{Z}_{p_0}(t,t_0| u)}{\delta u(\Vec{q})}
\\&+\alpha_s\int_z\int_{q^+\mathbf{q}} \mathcal{K}(z,q^+;t)\left[u(z \overrightarrow{q})u((1-z)\overrightarrow{q})-u(\overrightarrow{q})\right] \frac{\delta  \mathcal{Z}_{p_0}(t,t_0| u)}{\delta u(\Vec{q})} \, .
\end{split}
\end{equation}
Finally, the inclusive one-gluon distribution $D(x,\mathbf{k},t)$, which represents the probability of observing a gluon at time $t$ with momentum fraction $x$ and transverse momentum $\mathbf{k}$, can be obtained from the functional~\cite{BDIM2}
\begin{equation}\label{eq:final2}
D(x,\mathbf{k},t)\equiv k^+ \left(\frac{\delta \mathcal{Z}_{p_0}(t,t_0 | u)}{\delta u(\overrightarrow{k})}\right)_{u=1}    \, .
\end{equation}
Finally, using eq. \eqref{eq:Z_t_evolution_1} and noticing that $\cK(z,p_0^+;t)=\cK(1-z,p_0^+;t)$, we obtain
\small
\begin{equation}\label{eq:D_k_evolution}
\begin{split}
\partial_t D(x,\mathbf{k},t)&=\int_\mathbf{l} C(\mathbf{l},t) D(x,\mathbf{k}-\mathbf{l},t)  
\\&+\alpha_s \int_z \left[\frac{2}{z^2}\mathcal{K}\left(z,\frac{x}{z}p_0^+;t \right)D\left(\frac{x}{z},\frac{\mathbf{k}}{z};t\right)\Theta(z-x)-\mathcal{K}\left(z,xp_0^+;t\right)D(x,\mathbf{k},t)\right] \, .
\end{split}
\end{equation}
\normalsize
This is the well-known rate equation taking into account multiple soft medium-induced gluon production derived in~\cite{bottom_up_therm,BDIM2}. It has a very simple interpretation: the $\mathcal{O}(\alpha_s^0)$ term corresponds to the broadening in momentum space occurring between in-medium splittings; the first term at $\alpha_s$ order corresponds to the production of a gluon with energy fraction $x$ and momentum $\mathbf{k}$ from a parton of the same kinematics enhanced by a $\frac{1}{z}$ factor; and the last term corresponds to a gluon with momentum fraction $x$ and transverse momentum $\mathbf{k}$ being displaced to another energy and momentum mode via a splitting inside the medium, such that the creation and annihilation rates are balanced -- and thus probability is conserved.\par

\section{Computing the interference term in the soft regime}\label{sec:computing}

\begin{figure}[h!]
    \centering
    \includegraphics[scale=0.8]{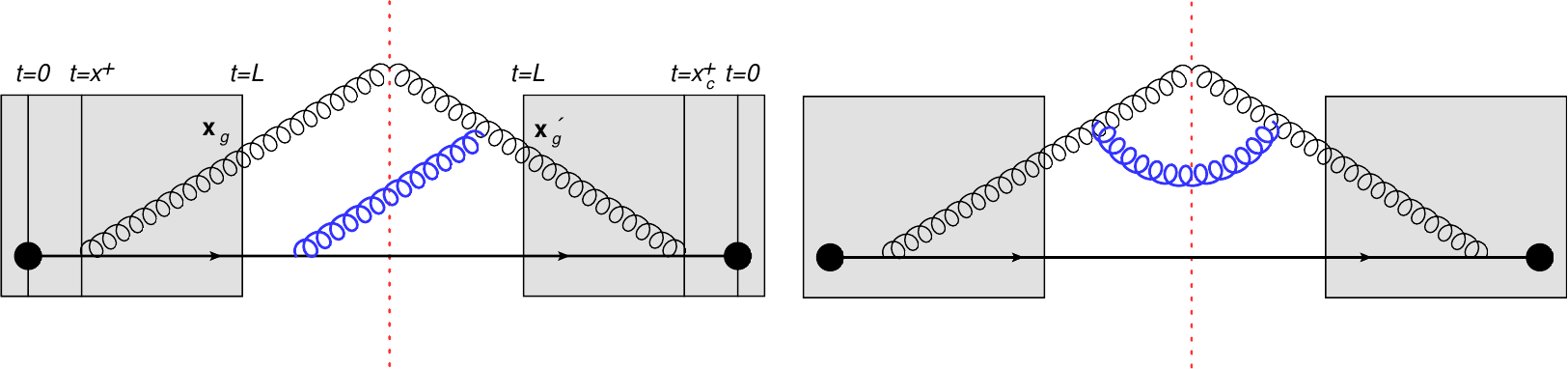}
    \caption{Left: The interference diagram $\mathcal{M}_{q}\mathcal{M}_g^\dagger$ computed in this paper with the soft gluon highlighted in blue. The remaining gluon leg is referred in the text as BDMPS-Z gluon. Right: The direct diagram $\mathcal{M}_{g}\mathcal{M}_g^\dagger$ that is not directly computed since it is proportional to the BDMPS-Z result. On the left hand side we have explicitly indicated all the time intervals present in the problem, with the emission time not matching in the amplitude ($x^+$) and its complex-conjugate ($x_c^+$). The medium is assumed to be static and homogeneous of length $L$, and the initiating quark is produced in a hard process (black blob) at time $t=t_0=0$ in both amplitude and complex-conjugate amplitude. Note that the initial hard process can be factorized from the rest of the diagram and is henceforth disregarded.}
    \label{fig:diagram1}
\end{figure}

In order to gauge the role of color coherence, we study the vacuum emission of a soft gluon after the single in-medium gluon emission considered in the previous sections. The process under consideration is depicted in Figure \ref{fig:diagram1}: the diagram in the right hand panel corresponds to the direct term, while the diagram on the left one corresponds to the interference contribution. Although both pieces have to be taken into account, the interference term is the only one that carries new physical information as it resolves the in-medium quark-gluon antenna. The equivalent BDMPS-Z contribution is obtained by removing the extra vacuum gluon from both diagrams. \par 
With that in mind, we start by computing the  eikonal emission amplitude of a gluon from either a quark or a gluon in vacuum, 
\begin{equation}
\mathcal{M}_q^{pure \, vac}=2gt^a\frac{\mathbf{k}^\prime \cdot \varepsilon_\perp^\prime}{\mathbf{k}^{\prime \ 2}}    \, ,
\end{equation}
\begin{equation}\label{eq:gg_emision_vac}
\mathcal{M}_g^{pure \, vac}=2igf^{abc}\frac{\mathbf{k}^\prime \cdot \varepsilon_\perp^\prime}{\mathbf{k}^{\prime \ 2}}\, .
\end{equation}
Using these results and the rules introduced in Appendix \ref{append:feynman_rules}, one can write the amplitudes for the case in which the two gluons are emitted either from the eikonal quark line ($\mathcal{M}_q$) or when the vacuum gluon comes from the gluon line ($\mathcal{M}_g$),
\begin{equation}
\mathcal{M}_q=-\frac{2g^2}{\omega}\int_{\mathbf{x}_g x^+}e^{-i\mathbf{k}\cdot\mathbf{x}_g}\partial_{\mathbf{x}}|_0    G^{ab}(L,x^+| \omega)\cdot \varepsilon_\perp t^c_{mi} U_{ij}(L,x^+) t^b_{jk} U_{kl}(x^+,0)  \frac{\mathbf{k}^\prime \cdot \varepsilon_\perp^\prime}{\mathbf{k}^{\prime^2}} \, ,
\end{equation}
\begin{equation}
\mathcal{M}_g=-\frac{2ig^2}{\omega}\int_{\mathbf{x}_g x^+}e^{-i\mathbf{k}\cdot\mathbf{x}_g}\partial_{\mathbf{x}}|_0    f^{cda}G^{ab}(L,x^+| \omega) \cdot \varepsilon_\perp U_{ij}(L,x^+) t^b_{jk} U_{kl}(x^+,0) \frac{\mathbf{k}^\prime \cdot \varepsilon_\perp^\prime}{\mathbf{k}^{\prime^2}} \, ,
\end{equation}
where $\mathbf{k}$ ($\omega$) and $\mathbf{k}^\prime$ ($\omega^\prime$) correspond to the transverse momentum (energy) of the in-medium BDMPS-Z gluon and the vacuum gluon, respectively. Their transverse polarization vectors are given by $\varepsilon_\perp$ and $\varepsilon_\perp^\prime$. We denote $\int_{x^+}\equiv\int_0^L dx^+$ and we have suppressed the dependence on transverse positions for clarity. We also implement the approximation in which the frequency of the vacuum gluon matches in amplitude and complex-conjugate amplitude. \par 
The interference amplitude $\mathcal{M}_{q}\mathcal{M}^\dagger_{g}$ is then given by
\small
\begin{equation}
\mathcal{M}_{q}\mathcal{M}^\dagger_{g}=-\frac{4 g^4i}{\omega^2(\mathbf{k}^\prime)^2}\int _{\mathbf{x}_g\mathbf{x}^\prime_g x^+ x_c^{ +}} e^{i  \mathbf{k}\cdot(\mathbf{x}_g^\prime-\mathbf{x}_g)}\partial_{\mathbf{x}}\cdot\partial_{\mathbf{x}^\prime}  \langle t^c_{mi}U_{ij}t^a_{jk}G^{ab}U_{kl}f^{dcb}U^\dagger_{l \overline{k}} G^{\dagger d \overline{a}} t^{\overline{a}}_{\overline{k}\overline{j}}U^{\dagger}_{\overline{j}m}   \rangle_{\mathbf{x}=\mathbf{x}^\prime=0}   \, ,  
\end{equation}
\normalsize
where $\langle \ \rangle $ indicates the medium average over all possible configurations of the background field -- see Appendix \ref{append:averages}. The color structure with the space-time arguments given explicitly and $U\equiv U(\mathbf{0})$ reads
\begin{equation}
\begin{split}
 &t^c_{ij} U_{jk}(L,x^+) t^a_{kl} G^{ab}(L,\mathbf{x}_g;x^+,\mathbf{x}|\omega)U_{lm}(x^+,0)
 \\\times&f^{dcb}U^\dagger_{m \overline{l}}(x^+_c,0) G^{\dagger d\overline{a}}(L,\mathbf{x}^\prime_g;x^+_c,\mathbf{x}^\prime|\omega) t^{\overline{a}}_{\overline{l}\overline{k}}U_{\overline{k}i}
^\dagger(L,x_c^+) \, ,
\end{split}
\end{equation}
where we have used that the vacuum emission is soft, so that the in-medium gluon has the same energy in amplitude and conjugate amplitude.
Combining the two middle Wilson lines and using the results shown in Appendix \ref{append:feynman_rules}, we obtain
\begin{equation}
\begin{split}
& \Tr({t^cU_{L,x^+}t^a U^\dagger_{x_c^+,x^+}t^{\overline{a}}U^\dagger_{L,x_c^+}}) G^{ab}(L,\mathbf{x}_g;x^+,\mathbf{x})f^{dcb}G^{\dagger d\overline{a}}(L,\mathbf{x}^\prime_g;x_c^+,\mathbf{x}^\prime) 
\\=&\Tr(U^\dagger_{L,x_c^+}t^cU_{L,x_c^+}t^{e}t^{\overline{a}})W^{\dagger e a}_{x_c^+,x^+}  G^{ab}(L,\mathbf{x}_g;x^+,\mathbf{x})f^{dcb}G^{\dagger d\overline{a}}(L,\mathbf{x}^\prime_g;x_c^+,\mathbf{x}^\prime) \, .
\end{split}
\end{equation}
Now using the well-known identity
\begin{equation}
t^a_{ij}t^b_{jk}=\frac{1}{2N_c}\delta^{ab}\delta_{ik}+\frac{1}{2}d^{abc}t^c_{ik}+\frac{i}{2}f^{abc}t^c_{ik}    \, ,
\end{equation}
and noticing that the only non-vanishing term has to be proportional to the $f$ symbol\footnote{ The singlet term vanishes as it is proportional to $\Tr(t)$, whereas the terms including the $d$ symbol vanish due to the allowed non-vanishing contractions between $d$ and $f$ symbols.}, the color structure simplifies to
\begin{equation}
\begin{split}
&\frac{i}{4} f^{e \overline{a} h} f^{dcb} W^{hc}_{L, x^+_c} W^{\dagger e a}_{x_c^+,x^+}G^{ab}(\mathbf{x}_g,\mathbf{x})_{L, x^+}G^{\dagger d \overline{a}}(\mathbf{x}^\prime_g,\mathbf{x}^\prime)_{L,x_c^+} 
\\=&\frac{i}{4}\int_{\mathbf{z}}f^{i jh}f^{dcb}W^{hc}_{L,x_c^+}W^{\dagger i a}_{x_c^+,x^+}  G^{al}(\mathbf{z},\mathbf{x})_{x_c^+,x^+}G^{lb}(\mathbf{x}_g,\mathbf{z})_{L,x_c^+}G^{\dagger{dj}}(\mathbf{x}^\prime_g,\mathbf{x}^\prime)_{L,x^+_c} \, ,
\end{split}
\end{equation}
where we have used the composition rule for Wilson lines -- see Appendix \ref{append:feynman_rules}. Finally, using the locality in light-cone time of the medium averages (see Appendix \ref{append:averages} or~\cite{BDIM1,Liliana1,Liliana2,Carlos_lectures}), we conclude the emission spectrum can be written as
\begin{equation}
 \begin{split}\label{eq:main_amplitude}
\omega \omega^\prime \frac{dI}{d^2\mathbf{k}d^2\mathbf{k}^\prime d\omega d\omega^\prime}&=-\frac{2C_FC_A\alpha_s^2}{(2\pi)^3\pi\omega^{2}N_c(N_c^2-1)^2 \mathbf{k}^{\prime 2}}\Re\Bigg[\int_{\mathbf{x}_g\mathbf{x}_g^\prime x^+ x^{+}_c \mathbf{z}}e^{i\mathbf{k}(\mathbf{x}^\prime_g-\mathbf{x}_g)}\partial_{\mathbf{x}}|_0\cdot \partial_{\mathbf{x}^\prime}|_0 
\\ &\times\Tr \langle G(\mathbf{z},\mathbf{x}| \omega) W^\dagger(\mathbf{0}))
\rangle_{x^+_c,x^+} 
\\ &\times
  f^{ijl}f^{abc}    \langle W^{ia}(\mathbf{0}) G^{jb}(\mathbf{x}_g,\mathbf{z}| \omega)G^{\dagger cl}(\mathbf{x}^\prime_g,\mathbf{x}^\prime| \omega) \rangle_{L,x_c^+}\Bigg] \, ,
\end{split}
\end{equation}
where we averaged over initial spin and color states, summed over the quantum numbers of the final states and used the shorthand $\int_{x^+x_c^+}\equiv \int_0^L dx^+ \int_{x^+}^L dx^+_c$. Note that the analogous BDMPS-Z spectrum is given by~\cite{Qw2}    
\begin{equation}
 \begin{split}\label{eq:BDMPS-ININ}
\left(\omega  \frac{dI}{d^2\mathbf{k}d\omega }\right)^{\mathbf{In-In}}&=\frac{2C_F\alpha_s}{(2\pi)^2\omega^{2}(N_c^2-1)^2}\Re\Bigg[\int_{\mathbf{x}_g\mathbf{x}_g^\prime x^+ x^{+}_c \mathbf{z}}e^{i\mathbf{k}(\mathbf{x}^\prime_g-\mathbf{x}_g)}\partial_{\mathbf{x}}|_0\cdot \partial_{\mathbf{x}^\prime}|_0 
\\ &\times\Tr \langle G(\mathbf{z},\mathbf{x}| \omega) W^\dagger(\mathbf{0}))
\rangle_{x^+_c,x^+}   \Tr\langle G(\mathbf{x}_g,\mathbf{z}| \omega)G^{\dagger}(\mathbf{x}^\prime_g,\mathbf{x}^\prime |\omega) \rangle_{L,x_c^+}\Bigg] \, .
\end{split}
\end{equation}
Additionally, we identify the vacuum gluon emission spectrum contribution
\begin{equation}
\left(\omega \frac{dI}{d\omega d^2\mathbf{k}}\right)^{\mathbf{g}}=\frac{\alpha_sC_A}{(2\pi^2)\mathbf{k}^2}  \, ,
\end{equation}
which can be easily obtained from eq. \eqref{eq:gg_emision_vac}. Combining it with eq. \eqref{eq:main_amplitude}, it results
\begin{equation}
 \begin{split}\label{eq:main_amplitude_2}
\omega \omega^\prime \frac{dI}{d^2\mathbf{k}d^2\mathbf{k}^\prime d\omega d\omega^\prime}&=-\left(\omega^\prime \frac{dI}{d\omega^\prime d^2\mathbf{k}^{\prime}}\right)^{\mathbf{g}}\frac{2C_F\alpha_s}{(2\pi)^2\omega^{2}N_c(N_c^2-1)^2 }\Re\Bigg[\int_{\mathbf{x}_g\mathbf{x}_g^\prime x^+ x^{+}_c \mathbf{z}}e^{i\mathbf{k}(\mathbf{x}^\prime_g-\mathbf{x}_g)}
\\ &\times \partial_{\mathbf{x}}|_0\cdot \partial_{\mathbf{x}^\prime}|_0 \Tr \langle G(\mathbf{z},\mathbf{x}| \omega) W^\dagger(\mathbf{0}))
\rangle_{x^+_c,x^+} 
\\ &\times
  f^{ijl}f^{abc}    \langle W^{ia}(\mathbf{0}) G^{jb}(\mathbf{x}_g,\mathbf{z}| \omega)G^{\dagger cl}(\mathbf{x}^\prime_g,\mathbf{x}^\prime| \omega) \rangle_{L,x_c^+}\Bigg] \, .
\end{split}
\end{equation}
For completeness, the remaining necessary steps for the full calculation of the spectrum are provided in Appendix \ref{append:averages}.

Comparing eqs. \eqref{eq:BDMPS-ININ} and \eqref{eq:main_amplitude_2}, we notice that, apart from the $\left(\omega^\prime \frac{dI}{d\omega^\prime d^2\mathbf{k}^\prime}\right)^\mathbf{g}$ factor, both results only differ in the color structure of the late-time medium average. Therefore, this piece must encapsulate the information related to the color coherence of the system. Under this observation, this will lead, in effect, to an \textit{ansatz} for the emission kernel entering the rate equation discussed in the previous section.

Within the harmonic oscillator approximation, the three point function in eq. \eqref{eq:main_amplitude_2} reads

\begin{equation}\label{eq:medium_average_new}
\begin{split}
&\int_\mathbf{z}^{\mathbf{x}_g} \mathcal{D}\mathbf{r}_1 \int_{\mathbf{x}^\prime}^{\mathbf{x}_g^\prime}\mathcal{D}\mathbf{r}_2 \exp \left(\int_t\frac{i\omega}{2} (\dot{\mathbf{r}}_1^2-\dot{\mathbf{r}}_2^2) -\frac{\hat{q}}{8 }\int_t\left(\mathbf{r}_2^2 +\mathbf{r}_1^2+(\mathbf{r}_2-\mathbf{r}_1)^2\right)\right)
\\=&\int_\mathbf{z}^{\mathbf{x}_g} \mathcal{D}\mathbf{r}_1 \int_{\mathbf{x}^\prime}^{\mathbf{x}_g^\prime}\mathcal{D}\mathbf{r}_2 \exp \left(\int_t\frac{i\omega}{2} (\dot{\mathbf{r}}_1^2-\dot{\mathbf{r}}_2^2) -\frac{\hat{q}}{4 }\int_t\left(\mathbf{r}_1\cdot\mathbf{r}_2+(\mathbf{r}_2-\mathbf{r}_1)^2\right)\right) \, ,
\end{split}    
\end{equation}
where $\mathbf{r}_1$ and $\mathbf{r}_2$ represent the transverse trajectories of the gluon in amplitude and its complex-conjugate, respectively. The term proportional to $(\mathbf{r}_1-\mathbf{r}_2)^2$ is the same one would obtain in the BDMPS-Z context.

As we wish to capture coherence effects between emitters at the level of the splitting without including final-state broadening effects, we follow the expansion of the in-medium propagator around the classical path introduced in~\cite{Altinoluk:2014oxa,Altinoluk:2015gia}, and write the full gluon propagator $G$ as a Wilson line $W$ evaluated along the classical path, i.e. $G(y^+,\textbf{y};x^+,\textbf{x}|\omega)\to G_0(y^+,\textbf{y};x^+,\textbf{x}|\omega)W(\textbf{x}_{\rm cl})$, with $\textbf{x}_{\rm cl}$ the classical path and $G_0$ the free propagator -- see Appendix \ref{append:feynman_rules}. The gluon propagators are effectively demoted to tilted Wilson lines in transverse space, with the trajectory fixed by the kinematics.

The trajectories $\mathbf{r}_{1,2}$ are expected to be straight lines diverging from the parent parton at an emission angle $\theta$ which will be set by the final kinematics. At large enough times, one can then expect that the $\mathbf{r}_1\cdot\mathbf{r}_2$ in the exponent in eq. \eqref{eq:medium_average_new} is dominated by the term quadratic in $t$:
\begin{equation}
\mathbf{r}_1(t)\cdot \mathbf{r}_2(t)\approx   \theta^2 t^2\, .
\end{equation}
The time integral for that term can then be performed exactly, thus yielding the usual simplified color coherence factor
\begin{equation}\label{eq:deltamed_sec3}
1-\Delta_{med}\equiv \exp\left(-\frac{\hat{q}}{12}\theta^2(L-x_c^+)^{3}\right)    \, ,
\end{equation}
in addition to the usual broadening experienced by the emitted gluon and encoded in the usual term proportional to $(\mathbf{r}_1-\mathbf{r}_2)^2$. Given that our focus is on including the proper factors which can capture the effects of color decoherence, the angle entering the formulas above is a function of the relative momentum of the outgoing antenna only, which is the momentum $\mathbf{Q}$ introduced in section \ref{subsec:BDMPS_full}.

The coherence factor in eq. \eqref{eq:deltamed_sec3} controls the contribution from the interference diagrams (see left panel of figure \ref{fig:diagram1}) and it leads to a simple physical interpretation in two regimes. The first corresponds to the case when either the emission angle or the in-medium propagation time after splitting, $L-x_c^+$, are sufficiently large, i.e. $\hat{q}L^3\theta^2\gg1$. In such cases $1-\Delta_{med}\approx 0$, and the contribution to the spectrum given by eq. \eqref{eq:main_amplitude_2} is vanishing. Heuristically, this limit corresponds to the case when the outgoing in-medium quark and gluon color fields become sufficiently randomized by the multiple interactions with the underlying medium, such that they effectively lose their color connection. As a consequence, there is no possibility of them exchanging color and thus the future interference for a soft vacuum gluon radiation is prohibited. On the other hand, a simple way to understand the case where the in-medium evolution is short or the antenna opening angle is small, $1-\Delta_{med}\approx 1+\mathcal{O}(\hat{q}\theta^2L^3)$, is to picture the transverse structure of the medium as an ensemble of color domains of size proportional to (the inverse) of the saturation scale $Q_s$, with different domains being color disconnected. Since the quark-gluon antenna transverse size is small, i.e. $Q_s^{-2}~\sim (\hat{q}L)^{-2}\gg \theta^2 (L-x_c^+)^2$, the system evolves in the medium with both outgoing quark and gluon states probing the same color domains, unable to break color coherence\footnote{The same picture is valid in the regime $1-\Delta_{med}\approx 0$. In this case, the antenna is large and thus the outgoing states always probe color uncorrelated domains in the medium, leading to the randomization of their color fields.}.

On top of the interference term we computed above explicitly, one has to consider the diagrams where the vacuum emission comes from the same parton in both the amplitude and conjugate amplitude (see right panel of figure \ref{fig:diagram1}). Due to the simple color structure, it is easy to observe this contribution to the spectrum leads to a term related to the vacuum emission of the soft gluon multiplying the in-medium contribution. Thus, in the two previous limiting regimes, the net result should be proportional to $1-(1-\Delta_{med})=\Delta_{med}$. In the regime where the outgoing states' lose color coherence, $\Delta_{med}=1$, one sees that the net result corresponds to the BDMPS-Z spectrum (up to a trivial vacuum term), as expected since this result assumes that states lose color coherence instantaneously after in-medium emission. On the other hand, when coherence is not lost, $\Delta_{med}\approx \mathcal{O}(\hat{q}\theta^2L^3)$, there is destructive interference between the two types of diagrams in figure \ref{fig:diagram1}, and the emission spectrum is suppressed. This picture is analogous to the one found for the in-medium QCD antenna~\cite{Antenna2}.

Beyond these limiting regimes and after considering both direct and interference diagrams, we see that then effect of considering color coherence can be implemented by including the coherence factor alongside the broadening after emission:
\begin{equation}\label{eq:correction}
\mathcal{P}_1(\mathbf{k};L,x_c^+)\to\mathcal{P}_1(\mathbf{k};L,x_c^+)\:\times\:\Delta_{med} \, .
\end{equation}

Even though this correction appears in the same region as the broadening post-emission, it makes sense instead to consider it as an additional factor to the emission kernel. A way to justify this observation is to note that the emission angle $\theta$ is controlled by the momentum and energy transfer during the almost instantaneous branching, as we will detail in the next section.
In fact, this \textit{ansatz} for the splitting kernel is qualitatively in agreement with the branching picture discussed in the above sections: branchings are still local in time and partons broaden independently, with the only difference that now there is a time delay before the medium resolves them. In addition, although we used a vacuum gluon to measure the coherence of the system, in Appendix~\ref{append:medium_emission} we argue that a similar structure to the last three point function in eq. \eqref{eq:main_amplitude_2} is still observed if one considers an in-medium gluon emission. 

This simple picture offers an opportunity to gauge the role of color coherence effects in the multiple soft emission regime, which is precisely the final goal of this paper.

\section{Introducing color coherence effects into the rate equation}\label{sec:evolution_eqs}
Our starting point is eq. \eqref{eq:P2_wQ}, which we repeat here in a slightly different form,

\begin{equation}\label{eq:P2_Q_mod}
\begin{split}
\cP_2(\mathbf{k},\mathbf{q},z;L,t_0)&=2g^2z(1-z)\int_{t_0}
^L dt \, \int_{\mathbf{m},\mathbf{Q}} \cK(\mathbf{Q},z,p_0^+;t)
\\
&\times\cP_1(\mathbf{m}-\mathbf{p}_0;t,t_0)\cP_1(\mathbf{k}-\mathbf{Q}-z\mathbf{m};L,t)\cP_1(\mathbf{q}+\mathbf{Q}-(1-z)\mathbf{m};L,t) \, .  
\end{split}
\end{equation}
In the BDMPS-Z framework, under the harmonic approximation~\cite{BDIM2}, the splitting kernel reads
\begin{equation}\label{eq:ck}
\begin{split}
\mathcal{K}(\mathbf{Q},z,p_0^+,t)
=\frac{2}{p_0^+}\frac{P_{gg}(z)}{z(1-z)}\sin\left(\frac{\mathbf{Q}^2}{2k_f^2(t)}\right)\exp\left(-\frac{\mathbf{Q}^2}{2k_f^2(t)}\right) \, ,
\end{split}
\end{equation}
where we have introduced the typical transverse momentum (squared) acquired due to in-medium scattering during the branching process, $k_f^2(t)=\sqrt{\hat{q}(t)(1-z(1-z))p_0^+(z(1-z))}$, with the time dependence vanishing in the plasma brick model. When the energy fraction $z$ is small, this gives the usual estimate $k_f^2\sim \sqrt{\hat{q}zp_0^+}$ implemented before. Integrating the previous equation over $\mathbf{Q}$ restores eq. \eqref{eq:K_energy} and justifies why $\mathbf{Q}^2\sim \sqrt{\hat{q}zp_0^+}\ll \hat{q}L$.\par
Following the discussion in the previous section, color coherence effects can be introduced in an effective way by the new splitting kernel 
\begin{equation}\label{eq:ckprime}
\begin{split}
&\cK^\prime(\mathbf{Q},z,p_0^+;L-t,t)\equiv\mathcal{K}(\mathbf{Q},z,p_0^+;t)\times\Delta_{med}(\mathbf{Q},z,p_0^+;L-t) \, ,
\end{split}
\end{equation}
where $\Delta_{med}$ is explicitly given by
\begin{equation}\label{eq:delta_med_sec4}
\Delta_{med}(\mathbf{Q},z,p_0^+;L-t)= 1- \exp \left( -\frac{\hat{q}}{12} \theta^2(\mathbf{Q},z,p_0^+)  (L-t)^3\right)  \, .
\end{equation}
Here the emission angle depends on the transverse momenta variables only through the relative transverse momentum between the two outgoing states $\mathbf{Q}$ and is fixed by the kinematics of the emission process, as considered in the previous section. It is explicitly given by~\cite{mapping_collinear}
\begin{equation}
\theta(\mathbf{Q},z,p_0^+)=\left|\frac{\mathbf{Q}}{z(1-z)p_0^+}\right| \, .
\end{equation}
In this approximation it is still true that $\mathbf{Q}^2\sim\sqrt{\hat{q}zp_0^+}$, and therefore $\mathbf{Q}^2\ll \mathbf{k}^2\sim\mathbf{q}^2\sim \hat{q}L$ still holds. Consequently, one can still neglect $\mathbf{Q}$ in the single particle broadening probabilities and integrate the new kernel $\cK^\prime$ over $\mathbf{Q}$, as presented in Section \ref{section:set_up}. In this way we recover the probabilistic picture of \cite{BDIM2} with the main difference that color coherence effects enter through the factor $\Delta_{med}$ in the kernel, which effectively delays the effect of the gluon emission thus accounting for the time it takes for the daughter parton to become a possible independent source of radiation. This modification incorporates into the evolution the fact that, for medium-induced emissions, the decoherence time might be larger than the formation time, as defined in \cite{mapping_collinear}.

The inclusion of the decoherence factor also implies that the emission kernel is no longer local in time, thus changing slightly the form of the evolution equations.
The new evolution equation for the (truncated) branching probability now reads
\small
\begin{equation}\label{eq:P2_final_1}
\begin{split}
&\partial_L\widetilde{\mathcal{P}}_2(\mathbf{k},\mathbf{q},z;L,t_0) = 2g^2z(1-z) \bigg\{\int_{\mathbf{Q}}  \mathcal{K}^\prime(\mathbf{Q},z,p_0^+;0,L) \delta(\mathbf{k}-\mathbf{Q}-z\mathbf{p}_0)\delta(\mathbf{q}+\mathbf{Q}-(1-z)\mathbf{p}_0)
\\&+\int_{t_0}^Ldt \, \int_{\mathbf{Q}}  \mathcal{K}(\mathbf{Q},z,p_0^+;t)\partial_L\Delta_{med}(z,\mathbf{Q},p_0^+;L-t) \delta(\mathbf{k}-\mathbf{Q}-z\mathbf{p}_0)
\delta(\mathbf{p}+\mathbf{Q}-(1-z)\mathbf{p}_0)\bigg\} \, .
\end{split} 
\end{equation}
\normalsize
The term in the first line vanishes since $\cK^\prime(\mathbf{Q},z,p_0^+;0,L)=0$ -- see eq.  \eqref{eq:delta_med_sec4}. This result can then be simplified to the form

\small
\begin{equation}\label{eq:evol_P2_sec4}
\begin{split}
\partial_L \widetilde{\mathcal{P}}_2(\mathbf{k},\mathbf{q},z;L,t_0)=2 g^2z(1-z)\int_{t_0}^L dt \, \Bar{\mathcal{K}}(z,p_0^+;L-t,t)(2\pi)^4\delta^{(2)}(\mathbf{k}-z\mathbf{p}_0)\delta^{(2)}(\mathbf{q}-(1-z)\mathbf{p}_0) \, ,
\end{split}
\end{equation}
\normalsize
with
\begin{equation}
\Bar{\mathcal{K}}(z,p_0^+;L-t,t)= \int_\mathbf{Q}\mathcal{K}(\mathbf{Q},z,p_0^+;t)\partial_L\Delta_{med}(\mathbf{Q},z,p_0^+;L-t)   \, .
\end{equation}
Since eq. \eqref{eq:evol_P2_sec4} has the same form as eq. \eqref{eq:evol_P2_sec1}, the generating functional evolution equation is similar to the one in eq. \eqref{eq:Z_t_evolution_1}. Therefore, one finds that the gluon distribution $D(x,\mathbf{k},t)$ obeys 
\begin{equation}\label{eq:D_k_evolution_sec4}
\begin{split}
&\partial_t D(x,\mathbf{k},t)=\int_\mathbf{l} C(\mathbf{l},t) D(x,\mathbf{k}-\mathbf{l},t)  
\\&+\alpha_s \int_0^t ds\,\int_z \left[\frac{2}{z^2}\Bar{\mathcal{K}}\left(z,\frac{x}{z}p_0^+;t-s,s \right)D\left(\frac{x}{z},\frac{\mathbf{k}}{z};s\right)\Theta(z-x)-\Bar{\mathcal{K}}\left(z,xp_0^+;t-s,s\right)D(x,\mathbf{k},s)\right] \, .
\end{split}
\end{equation}

The new modified kernel $\Bar{\mathcal{K}}$  has the effect of delaying the full effect of the emission over the decoherence time. At each step of the evolution in $t$ the effect of the emission at a previous time $s$ increases until the emitted gluon has completely decohered from its parent parton. This delay effect is particularly important to suppress further emissions, by not including the new gluon as a possible source of independent emissions until it is fully decohered from its parent parton.

\section{Conclusion and Outlook}\label{sec:conclusion}

In this paper we have implemented color coherence effects in a rate equation describing a full medium induced gluon shower. The obtained results are based on the calculation of the two gluon emission process depicted in Figure \ref{fig:diagram1}. 

In this work we have always considered the regime where multiple gluon emissions become important, i.e. when the gluon frequency is much smaller than the critical frequency, $\omega \ll \omega_c$, and consequently the gluon has a short formation time $t_f\ll L$. This leads to a Markovian/factorized picture, where in-medium branchings are well localized in time. In the fully decoherent picture previously considered in~\cite{BDIM2}, instantaneous branching is followed by final state broadening of the emitted gluons over a scale $\sim L$. 

 Going beyond this scenario, coherence effects are studied by introducing an extra soft vacuum gluon whose main role in the calculation is to allow measuring the degree of coherence of the outgoing quark-gluon antenna (see Figure \ref{fig:diagram1}). In the soft limit that we consider here, the propagation of the quark-gluon system after the formation time encodes both the broadening and the decoherence parameter entangled into a three point function. By making use of the tilted Wilson line approximation, we were able to factorize both contributions into the usual final state broadening and a coherence factor $\Delta_{ med}$ which dictates when the outgoing states are resolved individually by the medium. This coherence factor depends on an emission angle $\theta$, related to the momentum transfer $\textbf{Q}^2\ll \hat{q}L$ during the branching process and the associated energy fraction $z$. Since the momentum $\textbf{Q}^2$ acquired during branching is small compared to the momentum acquired due to broadening $\sim \hat{q}L$, at the level of the rate equations it can be ignored everywhere, as assumed in the original derivation of the rate equations in \cite{BDIM2}, except in the branching kernel $\cK(\textbf{Q},z)$. As such, when integrating over this momentum (or equivalently over the emission angle) to obtain the energy kernel $\cK(z)$, $\Delta_{ med}$ acts as an integration weight which suppresses the radiative contribution from highly coherent systems. In comparison, the incoherent case explored in~\cite{BDIM2}, assumes that the outgoing radiation decoheres instantaneously and hence, the integration weight is always unity -- see eq.~\eqref{eq:ck} compared to eq.~\eqref{eq:ckprime}.

Since the emission angle and the energy fraction $z$, which dictate the coherence between the outgoing states, are determined locally in the branching process, one can associate $\Delta_{med}$ to the splitting kernel $\cK$, with the length scale set by the broadening time. This split of the coherence factor and the broadening is of course a simplification we make so that the resumation of multiple emissions is possible. Nonetheless, in the future it would be relevant to incorporate broadening effects into the coherence factor, leading to a more realistic treatment of color coherence physics. 

The new evolution equation derived for the single gluon distribution, eq.~\eqref{eq:D_k_evolution_sec4}, becomes non-local in time, unlike previous totally decoherent results~\cite{BDIM2}. The non-locality of this equation relfects the fact that it takes a finite amount of time for a system to become resolved by the medium. In addition, the qualitative physical interpretation of the results, detailed in the main text, is in line with previous studies of color coherence effects in jet quenching~\cite{Antenna1,Antenna0,Antenna2,Antenna3,Antenna4}.

\acknowledgments{
The authors are grateful to N\'estor Armesto for helpful discussions. This work has received financial support from European Union's Horizon 2020 research and innovation program under the grant agreement No. 82409; from Xunta de Galicia (Centro singular de investigaci\'on de Galicia accreditation 2019-2022); from the European Union ERDF;  from the Spanish Research State Agency by “María de Maeztu” Units of Excellence program MDM-2016-0692 and project FPA2017-83814-P and from the European Research Council project ERC-2018-ADG-835105 YoctoLHC. The work of V.V. is supported by the Agence Nationale de la Recherche under the project ANR-16-CE31-0019-02. J.B. is supported by a fellowship from ``la Caixa" Foundation (ID 100010434) -- fellowship code  LCF/BQ/ DI18/11660057, and by funding from the European Union's Horizon 2020 research and innovation program under the Marie Sklodowska-Curie grant agreement No. 713673.}

\appendix
\section{Propagation of an energetic parton on a classical field}\label{append:feynman_rules}

When traversing a dense QCD medium, a parton with transverse momentum $\mathbf{p}$ and energy $p_0^+\gg |\mathbf{p}|$ keeps a straight trajectory in transverse space and only its color field gets rotated. In this high-energy regime, the in-medium propagator reduces to a Wilson line~\cite{Carlos_lectures,GYK}
\begin{equation}
W(\mathbf{x};L,t_0)=\mathcal{P}\exp\left(ig\int_{t_0}^L dx^+ \,A_-(\mathbf{x};x^+)\right)    \, ,
\end{equation}
so that the propagation is confined to a medium of length $L-t_0$. Here $A_-$ is the $-$ light-cone component of the classical background field describing the medium, while $\mathbf{x}$ is the transverse position at which the parton is located along the future light cone\footnote{We use two interchangeable notations for the light-cone time dependence: either it is given as an argument or it appears as a lower index.}. For simplicity, the gauge field's color indices are implicitly contracted with the generators of the color algebra. In the main text, we reserve the $W$ symbol for adjoint Wilson lines and $U$ for the fundamental representation case. In addition, we omit the transverse position as an argument when $\mathbf{x}=\mathbf{0}$. A useful relation between fundamental and adjoint Wilson lines is given by~\cite{Kovner,Liliana1,Liliana2,GYK,Carlos_lectures}
\begin{equation}
W^{\dagger ab}(\mathbf{x})=W^{ba}(\mathbf{x})= 2 \Tr[t^bU^\dagger(\mathbf{x}) t^aU(\mathbf{x})] \, ,
\end{equation}
where we have made the (adjoint) color indices explicit and traced in the fundamental representation. Two other useful identities are \par 
\begin{equation}
  \begin{split}
W^{ba}(\mathbf{x})t^a&=U(\mathbf{x})t^bU^\dagger(\mathbf{x})    \, , \\
t^bW^{ba}(\mathbf{x})&=U^\dagger(\mathbf{x})t^aU(\mathbf{x}) \, .
  \end{split} 
\end{equation}
Easing the eikonal restriction and including sub-eikonal corrections $\mathcal{O}(\mathbf{p}/p_0^+)$, yields the more general in-medium propagator~\cite{Carlos_lectures,Kovner,GYK}
\begin{equation}\label{eq:G_propagator}
G(\mathbf{x},L;\mathbf{y},t_0)=\int_\mathbf{y}^\mathbf{x} \mathcal{D}\mathbf{r} \, \exp\left( \frac{i p_0^+}{2}\int_{t_0}^L d\xi \, \dot{\mathbf{r}}^2(\xi)\right)\times W(L,t_0;\mathbf{r}(\xi)) \, ,
\end{equation}
where now the trajectory in transverse space is not fixed, with the eikonal propagator recovered in the limit $p_0^+\to \infty$. This propagator obeys the simple composition law
\begin{equation}
G(\mathbf{x},L;\mathbf{y},t_0)=\int_\mathbf{z} G(\mathbf{x},L;\mathbf{z},t)G(\mathbf{z},t;\mathbf{y},t_0) \, ,
\end{equation}
used in the main text in order to explore the $x^+$ locality of the medium averages.\par
Finally, we also make use of the results presented in~\cite{Altinoluk:2014oxa,Altinoluk:2015gia}, where a gradient expansion around the classical trajectory of $G(\mathbf{x},L;\mathbf{y},t_0)$ is performed. The leading result of such an expansion was referred to, in the main text, as the \textit{tilted} Wilson and it is given by
\begin{equation} \label{eq:bla1}
G(\mathbf{x},L;\mathbf{y},t_0)=G_0(\mathbf{x},L;\mathbf{y},t_0) W(\mathbf{x}_{\rm classical })_{L,t_0}+\mathcal{O}\left(\frac{L-t_0}{p_0^+}\partial^2_{\mathbf{x}_{\rm classical}}\right) \, .
\end{equation}
Here $G_0$ is the vacuum propagator (i.e. eq. \eqref{eq:G_propagator} with the gauge field term removed), and $\mathbf{x}_{\rm classical}$ is the classical trajectory in transverse space between positions $\mathbf{y}$ and $\mathbf{x}$ over the time interval $L-t_0$,
\begin{equation}
\mathbf{x}_{\rm classical}(s)=\mathbf{x}+\frac{\mathbf{x}-\mathbf{y}}{L-t_0}(s-L) \, , \quad t_0\leq s\leq L   \, .
\end{equation}
Note that eq. \eqref{eq:bla1} is derived under the assumption that $\frac{\mathbf{p}}{p_0^+}$ is finite.

\section{The medium averages}\label{append:averages}
This appendix outlines how to perform the medium averages present in our calculation. For a more detailed discussion on this topic, references such as \cite{GYK} should elucidate the reader.\par 
The basic object we wish to compute is the following two-point function in the adjoint representation\footnote{An analogous result can be derived in the fundamental representation.} within a time interval $L-t_0$,
\small
\begin{equation}
\frac{\Tr\langle W(\mathbf{x})W^\dagger(\mathbf{y})\rangle}{N_c^2-1}=   \frac{1}{N_c^2-1} \Tr\left\langle\exp\left(ig\int_{x^+} A_-(\mathbf{x};x^+)\right)\exp\left(-ig\int_{y^+} A_-^\dagger(\mathbf{y};y^+)\right) \right\rangle \, ,
\end{equation}
\normalsize
where we have used the explicit form of the Wilson line from Appendix \ref{append:feynman_rules}. In order to proceed, one expands in the coupling $g$ up to the first non-trivial order and models the field correlator as 
\begin{equation}\label{eq:medium_average_exp1}
\langle A_-^a(\mathbf{x};x^+)A_-^b(\mathbf{y};y^+)\rangle=\delta^{ab}\delta(x^+-y^+) B_\gamma(\mathbf{x},\mathbf{y};x^+) \, .
\end{equation}
Here $B_\gamma$ corresponds to the Fourier transform of the elastic in-medium scattering potential~\cite{Kovner:2001vi},
\begin{equation}
B_\gamma(\mathbf{x},\mathbf{y};x^+)=g^2n(x^+)\int_\mathbf{k} e^{i\mathbf{k}\cdot (\mathbf{x}-\mathbf{y})}\gamma(\mathbf{k})=   B_\gamma(\mathbf{x}-\mathbf{y};x^+)\, ,
\end{equation}
where $n(x^+)$ is the longitudinal density of scattering centers inside the medium and $\gamma$ the bare in-medium scattering potential, which in UV behaves as $\gamma(\mathbf{k})\sim 1/\mathbf{k}^4$. This connects to the dipole cross-section as follows,
\begin{equation}      
\sigma(\mathbf{x},t)=\sigma(\mathbf{x})n(t)= 2g^2\int_\mathbf{k}(1-e^{i\mathbf{k}\cdot \mathbf{x}})B_\gamma(\mathbf{x},t) \approx \frac{\hat{
q}(t)}{2C_A}\mathbf{x}^2 \log \frac{1}{\mu^2\mathbf{x}^2}\approx \frac{\hat{
q}(t)}{2C_A}\mathbf{x}^2 \log \frac{Q_c^2}{\mu^2} \, ,
\end{equation}
where in the next to last expression we have used the UV behavior of $\gamma$ and introduced $\hat{q}(t)=4\pi\alpha_s^2C_An(t)$. In the last expression we have also used the harmonic oscillator approximation and regulated the logarithm by introducing a large momentum scale $Q_c^2\sim 1/\mathbf{x}^2\gg\mu^2$, with $\mu$ the Debye mass.
Expanding the Wilson line to first non-trivial order produces
\begin{equation}\label{eq:medium_average_exp2}
W^{ab}(\mathbf{x})=\delta^{ab}+iA^c(\mathbf{x})f^{cab} -\frac{C_A}{2}\delta^{ab}B_\gamma(\mathbf{0})   \, .
\end{equation}
Under the above assumptions, the medium average at linear order, i.e. the leading contribution for $N=1$ in the opacity expansion, becomes
\begin{equation}
\frac{\Tr\langle W(\mathbf{x})W^\dagger(\mathbf{y})\rangle^{N=1}}{N_c^2-1}= 1-\frac{C_A}{2}\int_{x^+} n(x^+)\sigma(\mathbf{x}-\mathbf{y})= 1-\int_{x^+}\frac{\hat{q}(x^+)}{4} (\mathbf{x}-\mathbf{y})^2,
\end{equation}
with logarithmic contributions absorbed in $\hat{q}$. Re-exponentiating produces
\begin{equation}
\frac{\Tr\langle W(\mathbf{x})W^\dagger(\mathbf{y})\rangle}{N_c^2-1}= \exp\left(-\frac{C_A}{2}\int_{x^+}n(x^+)\sigma(\mathbf{x}-\mathbf{y}) \right)\equiv \mathcal{P}(\mathbf{x}-\mathbf{y};L,t_0)\, ,
\end{equation}
where we have introduced the dipole operator $\mathcal{P}(\mathbf{r})$, whose Fourier transform gives the single particle broadening probability, as already stated in Section \ref{section:set_up}.

 More complex averages can be computed following the same procedure. A relevant example, possible when sub-eikonal corrections are included, is the following two-point function, which under the harmonic oscillator approximation reads~\cite{GYK,Carlos_lectures}
\begin{equation}
\frac{\Tr \langle G(\mathbf{x},\mathbf{y})W^\dagger(\mathbf{0}) \rangle_{x_c^+,x^+}}{N_c^2-1}=\int_\mathbf{y}^\mathbf{x} \mathcal{D}\mathbf{r}\exp\left[ \frac{i\omega}{2}\int_{x^+}^{x_c^+} dt \,\dot{\mathbf{r}}^2-\int_{x^+}^{x_c^+}  dt \, \frac{\hat{q}(t)}{4}\mathbf{r}^2\right]    \, ,
\end{equation}
where on the left-hand side we have suppressed the dependence on the gluon frequency $\omega$. Here $G(\mathbf{x},\mathbf{y})$ is the full gluon propagator, and we have absorbed $\log Q_c^2/\mu^2$ in the definition of $\hat{q}$, as done in the previous example. Within the harmonic approximation implemented above, one can give an explicit solution~\cite{Liliana1,Liliana2}
\begin{equation}\label{eq:HO}
\begin{split}
&\frac{1}{N_c^2-1}\Tr\langle G(\mathbf{x},\mathbf{y})W^\dagger(\mathbf{0}) \rangle_{x_c^+,x^+}\equiv \cK(\mathbf{x},\mathbf{y})_{x_c^+,x^+} \\&=\frac{A(x_c^+,x^+)}{i\pi}\exp\bigg[iA(x_c^+,x^+)(B(x^+,x_c^+)\mathbf{x}^2+B(x_c^+,x^+)\mathbf{y}^2
-2 \mathbf{x}\cdot \mathbf{y})\bigg ]   \, .
\end{split}
\end{equation}
In particular, for a static and homogeneous medium the parametric functions $A$ and $B$ are given by
\begin{equation}
A(x_c^+,x^+)=\frac{\omega \Omega}{2\sin(\Omega (x_c^+-x^+))}\, , \quad B(x_c^+,x^+)=2\cos(\Omega(x_c^+-x^+))  \, ,
\end{equation}
while for other medium profiles we refer the reader to \cite{Arnold_simple_formula}. Here $\Omega$ is the harmonic oscillator frequency,
\begin{equation}
\Omega=\frac{1-i}{2}\sqrt{\frac{\hat{q}}{\omega}} \, .
\end{equation}

Another example is the three-point function appearing in eq. \eqref{eq:main_amplitude}. This correlator has already been explicitly computed in \cite{BDIM1} using the same techniques outlined above (see next appendix for some details). Explicitly, it can be written within the harmonic oscillator approximation as follows,

\begin{equation}
\begin{split}
&\int_\mathbf{z}^{\mathbf{x}_g} \mathcal{D}{\mathbf{r}_1}\int_{\mathbf{x}^\prime}^{\mathbf{x}_g^\prime} \mathcal{D}{\mathbf{r}_2}    \exp\left(\frac{i\omega}{2}\int_{x_c^+}^L dt \ \dot{\mathbf{r}}_1^2-\dot{\mathbf{r}}_2^2-\frac{\hat{q}}{8}\int_{x_c^+}^L dt \ \mathbf{r}_1^2+\mathbf{r}_2^2+(\mathbf{r}_1-\mathbf{r}_2)^2 \right)
\\=&\int_\mathbf{z}^{\mathbf{x}_g} \mathcal{D}{\mathbf{r}_1}\int_{\mathbf{x}^\prime}^{\mathbf{x}_g^\prime} \mathcal{D}{\mathbf{r}_2}   \exp\left(\frac{i\omega}{2}\int_{x_c^+}^L dt \ \dot{\mathbf{r}}_1^2-\dot{\mathbf{r}}_2^2-\frac{\hat{q}}{4}\int_{x_c^+}^L dt \ \mathbf{r}_1^2+\mathbf{r}_2^2-\mathbf{r}_1\cdot\mathbf{r}_2 \right) \, .
\end{split}
\end{equation}
When writing the dynamical terms we assumed that gluons' energies match in both amplitude and its complex-conjugate (otherwise the calculation is slightly more evolved, but still feasible). At this point, the unitary transformation 
\begin{equation}
\begin{split}
\mathbf{r}_1=\Gamma(\mathbf{r}_1^\prime+\beta \mathbf{r}_2^\prime) \, , \quad \mathbf{r}_2=\Gamma(\mathbf{r}_2^\prime+\beta \mathbf{r}_1^\prime) \, ,
\end{split}    
\end{equation}
is boost invariant and leaves the kinetic part of the Lagrangian unchanged, while the potential gives a diagonal term (with $ \Gamma^{-1}=\sqrt{1-\beta^2}$)
\begin{equation}
{\Gamma^2}\left[  \mathbf{r}_1^{ 2}(1-\beta+\beta^2)+\mathbf{r}_2^2(1-\beta+\beta^2)  \right]= \Gamma^2(1-\beta+\beta^2)(\mathbf{r}_1^2+\mathbf{r}_2^2) \, .
\end{equation}
Imposing that the cross-terms vanish results in 
\begin{equation}
1-4 \beta + \beta^2=0 \implies \beta = 2 \pm \sqrt{3} \, ,
\end{equation}
so that choosing the convergent solution $\beta=2-\sqrt{3}$, one gets 
\begin{equation}\label{eq:app2}
\begin{split}
   & \int_{\Gamma( \mathbf{z}-\beta \mathbf{x}^\prime)}^{\Gamma( \mathbf{x}_g-\beta \mathbf{x}_g^\prime)} D\mathbf{r}_1 \int_{\Gamma( \mathbf{x}^\prime-\beta \mathbf{z})}^{\Gamma( \mathbf{x}_g^\prime-\beta \mathbf{x}_g)} D\mathbf{r}_2 
    \exp\left( \int_{x_c^+}^{L}dt \frac{i\omega}{2}\dot{\mathbf{r}}^2_1   -\frac{\hat{q}\sqrt{3}}{8} \mathbf{r}_1^2 \right)
   \\ \times  & \exp\left(\int_{x_c^+}^{L}dt-\frac{i\omega}{2} \dot{\mathbf{r}}^2_2  -\frac{\hat{q}\sqrt{3}}{8} \mathbf{r}_2^2\right) \equiv \cJ(\mathbf{b}_g,\mathbf{b})_{L,x^+_c}\cJ^\dagger(\mathbf{c}_g,\mathbf{c})_{L,x_c^+} \, .
   \end{split}
\end{equation}
In this way, the present medium average is factorized into a product of the previous two-point correlator, with an effective diffusion coefficient $\hat{q}_{eff}=\hat{q}\frac{\sqrt{3}}{2}$. Note that we use $\cJ$ to denote $\cK$ with $\hat{q}\to \hat{q}_{eff}$, while the $\dagger$ denotes that one should use the conjugate harmonic oscillator frequency. Here we have also introduced
\begin{equation}
\mathbf{b}_g=\Gamma(\mathbf{x}_g-\beta \mathbf{x}_g^\prime) \, ,\quad \mathbf{b}=\Gamma(\mathbf{z}-\beta \mathbf{x}^\prime) \, ,\quad \mathbf{c}_g=\Gamma(\mathbf{x}_g^\prime-\beta \mathbf{x}_g) \, ,\quad \mathbf{c}=\Gamma(\mathbf{x}^\prime-\beta \mathbf{z})  \, . 
\end{equation}
Finally, these results can be used to simplify the spectrum in eq. \eqref{eq:main_amplitude} to the form
\begin{equation}\label{eq:full_solut,ion}
\begin{split}
 \omega \omega^\prime \frac{dI}{d^2\mathbf{k}d^2\mathbf{k}^\prime d\omega d\omega^\prime}&=-\left(\omega^\prime \frac{dI}{d\omega^\prime d^2\mathbf{k}^{\prime}}\right)^{\mathbf{g}}\frac{2C_F\alpha_s}{(2\pi)^2\omega^{2}}\Re\Bigg[\int_{\mathbf{x}_g\mathbf{x}_g^\prime x^+ x^{+}_c \mathbf{z}}e^{i\mathbf{k}(\mathbf{x}^\prime_g-\mathbf{x}_g)}
\\&\times \partial_{\mathbf{x}}\cdot \partial_{\mathbf{x}^\prime} 
  \cK(\mathbf{z},\mathbf{x})_{x_c^+,x^+}
   \cJ(\mathbf{b}_g,\mathbf{b})_{L,x_c^+} \cJ^\dagger(\mathbf{c}_g,\mathbf{c})_{L,x_c^+}\Bigg]_{\mathbf{x}=\mathbf{x}^\prime=\mathbf{0}} \, .
\end{split}
\end{equation}

\section{Outline of double in-medium gluon emission computation}\label{append:medium_emission} 
In this appendix, we study the interference diagram in the particular case where the second gluon emission happens inside the medium. We focus solely on the color structure of the squared amplitude, which can be directly read off from the diagram. The analysis of the full double in-medium gluon emission constitutes an extremely challenging problem in the literature (see~\cite{Iqbal1,Casalderrey_doublegg} for more in-depth discussions for both dense and dilute systems). In what follows, we will still assume that any formation time is much smaller than the medium length and we will not discuss the problem of overlapping formation times.

\begin{figure}[h!]
    \centering
    \includegraphics[scale=0.7]{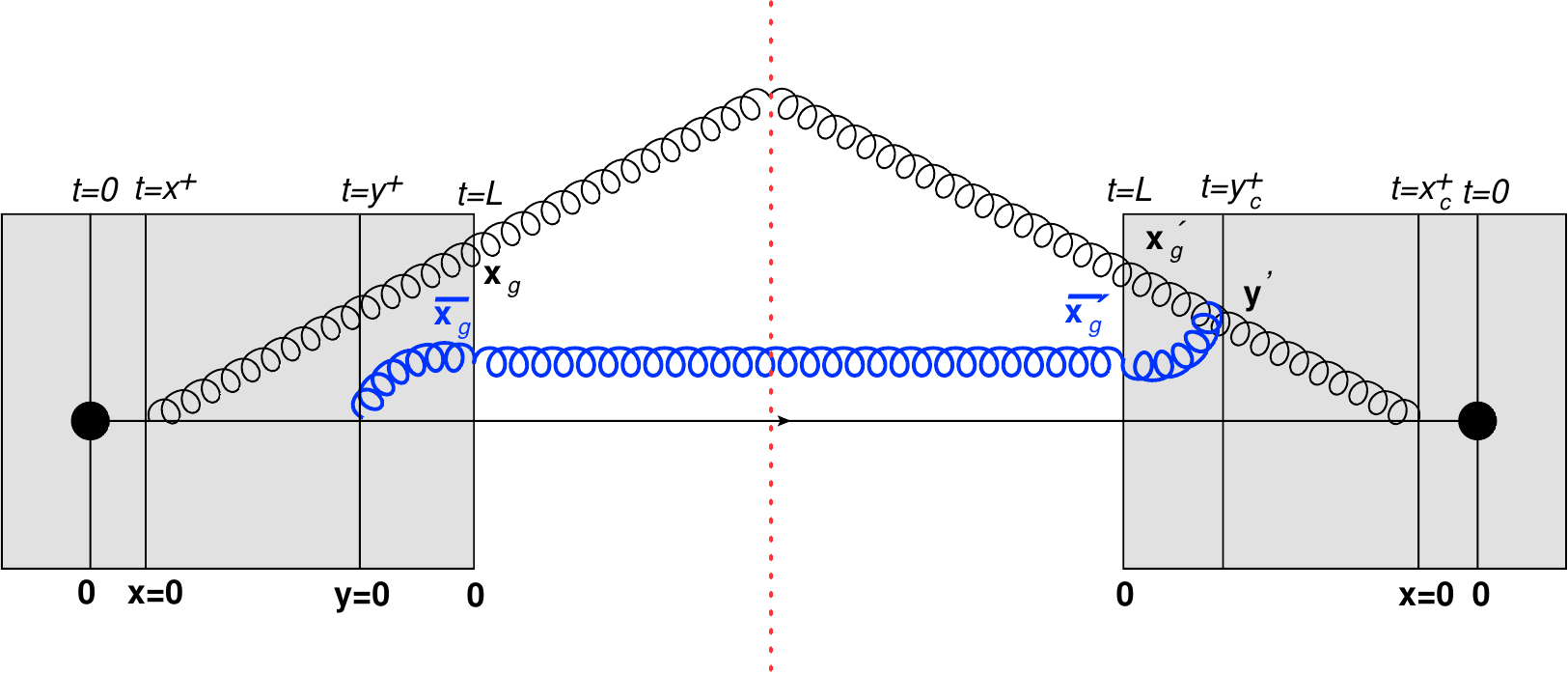}
    \caption{Leading color interference diagram for the case of double in-medium gluon emission. We identify the different time scales (above) and the transverse position of each splitting (below). In addition, we indicate the position in transverse space for the outgoing states.}
    \label{fig:med_in}
\end{figure}
In Figure \ref{fig:med_in}, we depict the leading color interference diagram associated to this process. As pointed out before, there are many ways one can order the time scales $x^+$, $x_c^+$, $y^+$ and $y_c^+$, but for the purposes of this appendix we will take $y^+_c>y^+>x_c^+>x^+$. We also use that the hard quark loses energy $\omega$ due to the gluon emission, with the first emission (in amplitude) having energy $\xi \omega$ and the second $(1-\xi)\omega$. In particular, the color structure of the depicted process reads 
\begin{equation}
\begin{split}
&\langle U_{ni}(L,y^+) t^c_{ij} G^{cd}(L,\Bar{\mathbf{x}}_g;y^+,\mathbf{0}|(1-\xi)\omega)U_{jk}(y^+,x^+)t^{a}_{kl}U_{lm}(x^+,0)G^{ab}(L,\mathbf{x}_g;x^+,\mathbf{0}|\xi\omega)
\\&\times G^{\dagger b\Bar{b}}(L,\mathbf{x}_g^\prime;y_c^+,\mathbf{y}^\prime|\xi\omega)G ^{\dagger d\Bar{d}}(L,\Bar{\mathbf{x}}_g^\prime;y_c^+,\mathbf{y}^\prime|(1-\xi)\omega)f^{\Bar{c}\Bar{d}\Bar{b}}G^{\dagger\Bar{c}\Bar{a}}(y_c^+,\mathbf{y}^\prime;x_c^+,\mathbf{0}|\omega)
\\& \times U^{\dagger}_{m\Bar{l}}(x_c^+,0)t^{\Bar{a}}_{\Bar{l}\Bar{k}} U^{\dagger}_{\Bar{k}n}(L,x_c^+)\rangle    \, ,
\end{split}    
\end{equation}
where eikonality is assumed for the leading parton, such that $\mathbf{x}=\mathbf{y}\equiv\mathbf{0}$\footnote{In the triple gluon vertex at $\mathbf{y}^\prime$ one would need to introduce dummy variables in order to evaluate the derivative operators appearing in the full squared amplitude, and only then set all positions to $\mathbf{y}^\prime$. However, for the current discussion this subtlety does not become relevant since we are only interested in the color structure.}. After some algebraic manipulations and assuming $\xi\to1$ as done in the main text, such that $(1-\xi)\omega\to \omega^\prime$ and $\xi\omega\to \omega$, it results

\begin{equation}
\begin{split}
&\int_{\mathbf{z}_1\mathbf{z}_2\mathbf{z}_3\mathbf{z}_4\mathbf{z}_5}  \Tr \langle G(\mathbf{z}_1;\mathbf{0}|\omega)W^\dagger(\mathbf{0})\rangle_{x_c^+,x^+}\langle f^{l\Bar{a}h} W^{h c}(y^+,x_c^+)G^{c\alpha}(y_c^+,\mathbf{z}_2;y^+,\mathbf{0}|\omega^\prime)G^{\alpha d}(L,\Bar{\mathbf{x}}_g;y_c^+,\mathbf{z}_2|\omega^\prime)
\\&\times G^{\dagger d \Bar{d}}(L,\Bar{\mathbf{x}}_g^\prime;y_c^+,\mathbf{y}^\prime|\omega^\prime)G^{\dagger b \Bar{b}}(L,\mathbf{x}_g^\prime;y^+_c,\mathbf{y}^\prime|\omega)G^{l\beta}(y^+,\mathbf{z}_3;x_c^+,\mathbf{z}_1|\omega)G^{\beta \sigma}(y_c^+,\mathbf{z}_4;y^+,\mathbf{z}_3|\omega)
\\&\times G^{\sigma b}(L,\mathbf{x}_g;y_c^+,\mathbf{z}_4|\omega)f^{\Bar{c}\Bar{d}\Bar{b}}G^{\dagger\Bar{c}\gamma}(y^+_c,\mathbf{y}^\prime;y^+,\mathbf{z}_5|\omega)G^{\dagger\gamma\Bar{a}}(y^+,\mathbf{z}_5;x_c^+,\mathbf{0}|\omega)\rangle  \, .
\end{split}    
\end{equation}
It is now possible to break down the medium averages for each time interval as follows,
\begin{equation}
\begin{cases}
f^{l\Bar{a}h} \langle W^{hc}G^{l\beta}G^{\dagger\gamma\Bar{a}}\rangle   \qquad & {\rm in} \, (x_c^+,y^+),
\\  f^{\Bar{c}\Bar{d}\Bar{b}}\langle G^{c\alpha}G^{\beta \sigma} G^{\dagger  \Bar{c}\gamma}\rangle   \qquad &{\rm in} \, (y^+,y_c^+),
\\  \langle G^{\alpha d}G^{\dagger d\Bar{d}}G^{\sigma b} G^{\dagger b\Bar{b}}\rangle  \qquad &{\rm in} \, (y_c^+,L) \, .
\end{cases}    
\end{equation}
The color structure of this system is quite involved and cannot be written in a closed form~\cite{BDIM1, Liliana2}. In order to proceed, we take again the late time evolution of the system to be that of two independent color dipoles, so that we obtain
\begin{equation}\label{eq:lll}
\begin{cases}
 f^{lah} \langle W^{h c}G^{l\beta}G^{\dagger  \gamma a } \rangle \qquad & {\rm in} \, (x_c^+,y^+),
\\f^{idb}\langle G^{cd}G^{\beta b} G^{\dagger i\gamma }\rangle \qquad &{\rm in}\, (y^+,y_c^+),
\\ \Tr \langle G(\Bar{\mathbf{x}}_g;\mathbf{z}_2|\omega^\prime)G^{\dagger}(\bar{\mathbf{x}}_g^\prime;\mathbf{y}^\prime|\omega^\prime)\rangle \Tr \langle G(\mathbf{x}_g;\mathbf{z}_4|\omega) G^\dagger(\mathbf{x}_g^\prime;\mathbf{y}^\prime| \omega)\rangle \qquad& {\rm in} \, (y_c^+,L) \, .
\end{cases}    
\end{equation}
For the intermediate medium average, we extract the overall color factor by making use of the techniques introduced in the previous appendix. Taking eqs. \eqref{eq:medium_average_exp1} and \eqref{eq:medium_average_exp2}, we consider the contribution of an extra scattering center to the medium average at time $\tau$, such that the time interval is split into two sub-intervals: the former from $(y^+,y_c^+-\tau)$, while the latter (infinitesimal one) is $(y_c^+-\tau,y_c^+)$. Thus, we can write any Wilson line as

\begin{equation}\label{eq:Wilson_exp}
W^{ij}(\mathbf{x})_{y^+,y_c^+}=   W^{ik}(\mathbf{x})_{y_c^+-\tau,y^+}\left(\delta^{kj}\left(1-\frac{C_A}{2}B_\gamma(\mathbf{0})\right)-if^{ksj}A^s(\mathbf{x})\right)_{y_c^+,y_c^+-\tau} \, .
\end{equation}
Finally, by using eq. \eqref{eq:Wilson_exp} in the second line of eq. \eqref{eq:lll} and after the some color algebra we obtain, in accordance with the results shown in the main text, 
\begin{equation}\label{eq:mmm}
\begin{split}
 &f^{idb}\langle W^{cd}(\mathbf{x}_1)W^{\beta b}(\mathbf{x}_2)W^{\dagger i\gamma}(\mathbf{x}_3)\rangle_{y^+,y_c^+}\sim    f^{idb} \langle W^{cd}(\mathbf{x}_1)W^{\beta b}(\mathbf{x}_2)W^{\dagger i\gamma}(\mathbf{x}_3)\rangle_{y_c^+-\tau,y^+}
 \\&\times\left(1-\frac{3C_A}{2}B_\gamma(\mathbf{0})+\frac{C_A}{2}[B_\gamma(\mathbf{x}_1-\mathbf{x}_2)+B_\gamma(\mathbf{x}_3-\mathbf{x}_2)+B_\gamma(\mathbf{x}_1-\mathbf{x}_3)]\right)_{y_c^+,y_c^+-\tau} \, .
 \end{split}
\end{equation}
The time locality of the medium averages and this result leads to the conclusion that the medium average within the time interval $(x_c^+,y^+)$ must take the form 
\begin{equation}\label{eq:sss}
\begin{split}
& f^{hla}f^{ijk}\langle W^{hi}G^{lj}G^{\dagger k a}\rangle_{x_c^+,y^+} \, ,
\end{split}    
\end{equation}
matching the result in the main text. For other orderings of the branching times in amplitude and conjugate amplitude, we find the same result for the earliest time interval.\par

\bibliographystyle{JHEP.bst} 
\bibliography{Lib.bib}

\end{document}